\documentclass[12pt]{article}
\usepackage{amsmath}
\usepackage{graphicx,psfrag,epsf}
\usepackage{enumerate}
\usepackage{textcomp}
\usepackage{setspace}
\usepackage[margin=1in]{geometry}

\usepackage{natbib}
\usepackage{float}
\usepackage{caption}
\usepackage{hyperref}

\newcommand{\amend}[1]{{\color{black} #1}}

\hypersetup{
  colorlinks=true,
  linkcolor={magenta},
  citecolor={magenta},
  urlcolor={magenta}
}

\title{Fractional Tackles: Leveraging Player Tracking Data for Within-Play
Tackling Evaluation in American Football}
\author{
Quang Nguyen\(^\text{\scriptsize 1}\) 
\quad Ruitong Jiang\(^\text{\scriptsize 2}\) 
\quad Meg Ellingwood\(^\text{\scriptsize 1}\) 
\quad Ronald Yurko\(^\text{\scriptsize 1}\)
}
\date{
\(^\text{\scriptsize 1}\)Department of Statistics \& Data Science\\
\(^\text{\scriptsize 2}\)Neuroscience Institute and Center for the Neural Basis of Cognition\\
Carnegie Mellon University\\
Pittsburgh, PA 15213\\
}

\begin{document}
\maketitle
\begin{abstract}
Tackling is a fundamental defensive move in American football, with the main purpose of stopping the forward motion of the ball-carrier.
However, current tackling metrics are manually recorded outcomes that are inherently flawed due to their discrete and subjective nature.
Using player tracking data, we present a novel framework for assessing tackling contribution in a continuous and objective manner. 
Our approach first identifies when a defender is in a ``contact window'' of the ball-carrier during a play, before assigning value to each window and the players involved. 
This enables us to devise a new metric called fractional tackles, which credits defenders for halting the ball-carrier's forward motion toward the end zone.
We demonstrate that fractional tackles overcome the shortcomings of traditional metrics such as tackles and assists, by providing greater variation and measurable information for players lacking recorded statistics like defensive linemen.
We view our contribution as a significant step forward in measuring defensive performance in American football and a clear demonstration of the capabilities of player tracking data.
\end{abstract}

\noindent%

\newpage

\setstretch{1.45}

\section{Introduction}
\label{sec:introduction}

American football is a complex sport where a game evolves continuously in time, both physically and tactically.
Within a play, there are numerous elements that interact with each other to form a dynamic and complex system. 
Tackling is an important aspect in American football.
Once the ball is successfully handed off to a running back or thrown to a receiver, the main task of every defender on the field is to tackle the ball-carrier and prevent them from gaining yardage on offense.
The most basic tackling statistics are tackles, either solo or assisted, which are observed measures of tackling outcome at the play level.
Solo tackles credit a defender when they are the primary player to tackle the ball-carrier, whereas assisted tackles (or assists) are recorded when two or more defenders make a tackle at the same time.
Additionally, the sum of solo and assisted tackles are commonly referred to as combined tackles.
Surprisingly, tackles are not an official statistic in the National Football League (NFL).
During each game, the home team score keeper is tasked with tracking whether a player records a tackle.
Although tackles are frequently used across various forms of football media coverage, many over the years have questioned its accuracy \citep{monson2012nfl}, with phrases like ``make-believe stat'' \citep{clark2012nfl} and ``blurry, elusive statistic'' \citep{mcknight2015farce} being used to describe the shortcomings of tackles.

There exist several attempts to improve the binary box-score tackles metrics.
\cite{burke2010tackle} proposes ``tackle factor'' which describes the percentage of a defender's total tackles relative to the number of tackles accumulated by their team over a certain time span (e.g., season).
StatsBomb defines a metric called ``true tackles'' by treating an assist as 0.5 when combining solo and assisted tackles \citep{bursik2023enhancing}.
Pro Football Focus (PFF) offers ``stops'' which indicates whether a defender records a tackle on plays where the offense does not gain positive yardage \citep{jahnke2019defensive}.
PFF also tracks ``missed tackles'' as an indicator for whether a defender misses a tackle on a given play \citep{mellor2019signature}.
These metrics are simple; yet limited due to their discrete and subjective nature, similar to the limitations of the traditional tackling counts.
\amend{
The central challenge here is data quality, as data recorded at the play level only capture discrete outcomes and reductive summary statistics (e.g., tackles) rather than more granular information like continuous-time within-play player movement.
Hence, it is difficult to devise any meaningful metric that properly credits defenders for their tackling contribution using data available in a low level of resolution.
}

Fortunately, the advent of player tracking data provides an exceptional opportunity to develop novel and actionable approaches for evaluating player performance in sports \citep{baumer2023big, kovalchik2023player}.
This fine-grained resource captures information about the positioning and movement of every athlete on the playing surface at a high frequency.
In American football, the NFL collects tracking data via the Next Gen Stats (NGS) tracking system \citep{nfl2024ngs} by embedding radio frequency identification (RFID) tags in player shoulder pads and also inside the football.
This returns two-dimensional locations of every player on the field and the ball at a rate of ten measurements per second, yielding over one hundred million space-time observations over the course of a regular season.
From here, other tracking attributes such as distance traveled, speed, and orientation can be straightforwardly computed \citep{lopez2020bigger}.
NGS tracking data have become a dynamic part of live NFL games, providing real-time insights into player and team performance.
For instance, information such as ball-carrier's top speed, pass coverage matchup, and expected rushing yards are often presented during nationally televised NFL games in America (e.g., Sunday Night Football on NBC).
Recently, Amazon has introduced an alternate Thursday Night Football broadcast known as Prime Vision, aiming to enhance the NFL viewing experience with statistics and technology powered by Amazon Web Services and Next Gen Stats.

In addition, to crowdsource public insight into tracking data, the NFL hosts an annual event called the Big Data Bowl \citep{lopez2020bigger}. 
Since its first edition in 2019, the NFL Big Data Bowl has released yearly samples of player tracking data to go along with football-specific competition themes such as pass coverage, special teams, linemen, and tackling.
These efforts have led to recent data science innovations in football, including within-play expected points \citep{yurko2020going}, receiver route identification \citep{chu2020route}, pass coverage annotation \citep{dutta2020unsupervised}, and pass rush pressure evaluation \citep{nguyen2024here}.
It is noteworthy that some of the aforementioned contributions analyze aspects on defense in football such as pass coverage and pass rush.
These are significant breakthroughs, as defensive performance has been an understudied part of football, especially relative to aspects on offense, most notably quarterback evaluation \citep{burke2019deepqb, deshpande2020expected, reyers2023quarterback}.
Now, player tracking data provide ample opportunities for novel
advancements in evaluating American football defenders.

In this work, we rely on the high-resolution spatiotemporal tracking data to study tackling---a crucial defensive move in American football---at the continuous-time within-play level.
Specifically, we introduce a novel framework for assigning defensive credit for tackling contribution throughout a play.
To develop our framework, we first need to think carefully about the purpose of a tackle in football.
According to the NFL Rulebook \citep{nfl2023rulebook},
\begin{quote}
Tackling is an attempt by a defensive player to hold a runner to halt his advance or bring him to the ground. \hfill (Section 35, NFL Rulebook)
\end{quote}
Thus, the conceptualization of our framework is based on this definition, as we focus on measuring defensive contribution in terms of halting the forward motion of the ball-carrier.
This forward motion can be measured via the ball-carrier's velocity in the target end zone direction, which can be directly-derived from the observed tracking data. 
Our framework consists of the following steps.
First, rather than relying solely on observed tackle events, we leverage the tracking data to identify when a defender is in a contact window of the ball-carrier.
We then measure the change in the ball-carrier's velocity during each window, and use this to divide credit between the defenders for stopping the ball-carrier's velocity toward the target end zone.
This allows us to create a model-free, continuous metric for assessing defensive contributions within a play that moves beyond the binary tackle designation.

The remainder of this paper is organized as follows.
In Section \ref{sec:methods}, we discuss the player tracking data and detail our methodology.
Next, in Section \ref{sec:results}, we present applications and assessments our proposed framework.
Finally, in Section \ref{sec:discussion}, we discuss our findings, limitations of this study, and potential future work.

\section{Methods}
\label{sec:methods}

\subsection{Player tracking data}
\label{sec:tracking}

For our analysis, we use player tracking data provided by the NFL Big Data Bowl 2024 \citep{lopez2023nfl}, which has the competition theme of tackling.
The data span 12,486 plays across 136 games during the first nine weeks of the 2022 NFL season.
Most importantly, the data provide tracking information for all 22 players on the field (and the football), including a player's spatial coordinates, speed, acceleration, angle of motion, and orientation recorded at a rate of 10 Hz (i.e., 10 data points per second).
In addition, event labels (e.g., ball snap, handoff, first contact, tackle, etc.) are provided for specific frames within each play in the data.
Later on, we use the first contact and tackle annotated events to identify whether a defender is in a contact window of the ball-carrier in Section \ref{sec:framework}.
Table \ref{tab:tracking} shows a tracking data example for a \href{https://youtu.be/_78KVrbDA2g&t=34}{seven-yard run by New York Giants running back Saquon Barkley} against the Tennessee Titans during week 1 of the 2022 NFL regular season.
In this play, the annotated events are ball snap, handoff, first contact, and tackle

To supplement the player tracking data, the NFL Big Data Bowl also provides data on player, play, and game-level characteristics, each of which can be matched with the tracking data via a unique identifier.
Since the focus is on tackling, information regarding whether a player records a tackle or assist, forces a fumble (all provided by the NFL) and misses a tackle (tracked by Pro Football Focus) for each play are also available.
Note that additional contextual information can be obtained by merging the provided data with publicly-available play-by-play data resources, such as the \texttt{nflfastR} \texttt{R} package \citep{carl2024nflfastr, r2024language}.

\begin{figure}[!h]
\centering
\includegraphics[width=0.7\linewidth]{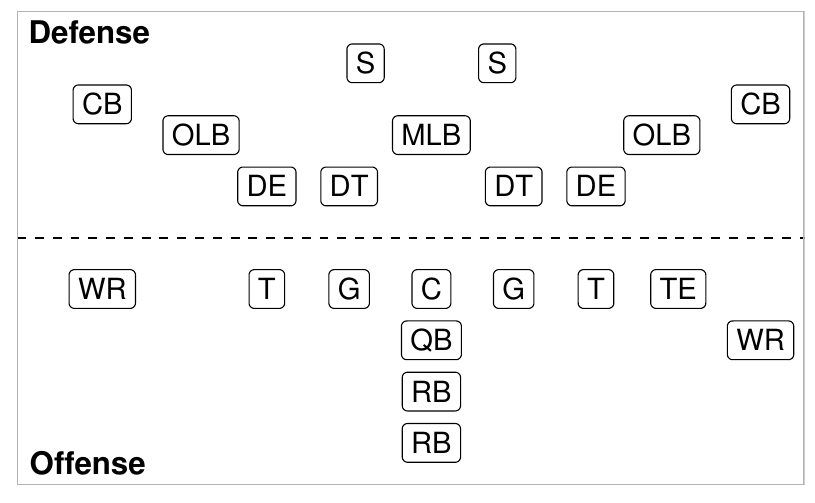}
\caption{
\amend{
An example schematic of American football positions.
The dashed line depicts the line of scrimmage separating the defense and offense.
\textit{Top (on defense):}
Defensive tackles (DT) and defensive ends (DE) are part of the defensive line and often attempt to stop the ball-carrier as early as possible after the ball is handed off on run plays.
Outside and middle linebackers (OLB and MLB) play behind the defensive line and involve in defending against run plays.
Cornerbacks (CB) and safeties (S) are usually in charge of covering opposing receivers on pass plays.
\textit{Bottom (on offense):}
Tackles (T), guards (G) and center (C) form an offensive line to block the opposing defensive line players and protect the quarterback (QB), with the center also responsible for snapping the ball. 
The QB, whose primary role is to lead the offense and direct the team's play on the field, usually handles the ball from the center to start the play.
Running backs (RB) typically line up behind the offensive line and QB, with one back (halfback) responsible for carrying the ball after receiving a QB handoff on run plays while the other (fullback) focusing more on blocking.
Wide receivers (WR) have the main role of running routes and catching passes.
Tight ends (TE) usually line up next to the tackles and play as either a blocker or receiver depending on the play.
Here, the defense employs what known as the \textit{4-3 formation}, whereas the offense lines up in the so-called \textit{I formation}.
}
}
\label{fig:pos_diagram}
\end{figure}

\amend{
Our analysis focuses on all run plays by running backs (RBs) from the first nine weeks of the 2022 NFL regular season.
For context, the primary duty of a RB in American football on rushing plays is to carry the ball after receiving a handoff from the quarterback (see Figure \ref{fig:pos_diagram} for more information on all American football positions).
In doing so, the RB often contributes by evading tackles using their strength and agility in order to advance the ball on the ground and gain extra yardage for the offense.
Across the provided nine-week data sample, there are 6,670 total run plays.
Since the majority of ball-carriers on these plays are RBs, we do not consider rarer scenarios such as quarterback rushes, wide receiver jet sweeps, or tight end direct snaps. 
This results in 5,539 total RB run plays for our forthcoming analysis.
Here, to identify the considered run plays, we use the play type attribute provided by \texttt{nflfastR} and extract plays where the ball-carrier's position is RB.
Furthermore, we rely on the event annotation in the tracking data to identify the ball-carrier sequence for each play.}
In particular, we consider all frames between the ball snap and the end-of-play event (e.g., tackle, out-of-bound, touchdown, etc.).
We also standardize player coordinates and directions to ensure that the offensive teams are moving in the same direction (left to right) in every play. 

\begin{table}[t]
\caption{Example of tracking data for a play during the New York Giants versus Tennessee Titans NFL game on September 11, 2022. The data presented here are for Giants running back Saquon Barkley, and the frames included are between the ball snap and when the tackle is made. The data attributes include frame identifier (frameId), two-dimensional coordinates (x and y), speed (s), acceleration (a), distance traveled from previous frame (dis), orientation (o), direction (dir), and event tag for each frame (event). \label{tab:tracking}}
\centering
\begin{tabular}{rrrrrrrrl}
\hline
frameId & x & y & s & a & dis & o & dir & event \\ 
\hline
6 & 93.30 & 29.84 & 0.00 & 0.00 & 0.01 & 265.57 & 287.75 & ball\_snap \\ 
7 & 93.30 & 29.84 & 0.00 & 0.00 & 0.00 & 265.57 & 286.42 & -- \\ 
\vdots & \vdots & \vdots & \vdots & \vdots & \vdots & \vdots & \vdots & \vdots \\
19 & 93.46 & 27.03 & 5.29 & 3.37 & 0.51 & 247.01 & 180.15 & handoff \\ 
\vdots & \vdots & \vdots & \vdots & \vdots & \vdots & \vdots & \vdots & \vdots \\
49 & 83.41 & 8.73 & 5.17 & 4.15 & 0.53 & 232.38 & 249.56 & first\_contact \\ 
\vdots & \vdots & \vdots & \vdots & \vdots & \vdots & \vdots & \vdots & \vdots \\
63 & 78.83 & 7.61 & 1.35 & 3.01 & 0.15 & 174.70 & 226.38 & -- \\ 
64 & 78.74 & 7.51 & 1.24 & 2.55 & 0.13 & 172.53 & 214.98 & tackle \\ 
\hline
\end{tabular}
\end{table}

\subsection{Framework}
\label{sec:framework}

We now describe our framework for within-play tackling evaluation of NFL defenders. 
There are three main steps to our framework.
First, we identify contact windows, which are time intervals within a play where a defender is possibly making contact with the ball-carrier.
Next, we assign a value to each contact window based on the change in the ball-carrier's velocity toward the end zone during the window. 
Finally, we attribute the contact window value across all defenders involved and obtain our fractional tackles metric.

\subsubsection{Identifying contact windows}

\begin{figure}[t]
\centering
\includegraphics[width=0.85\linewidth]{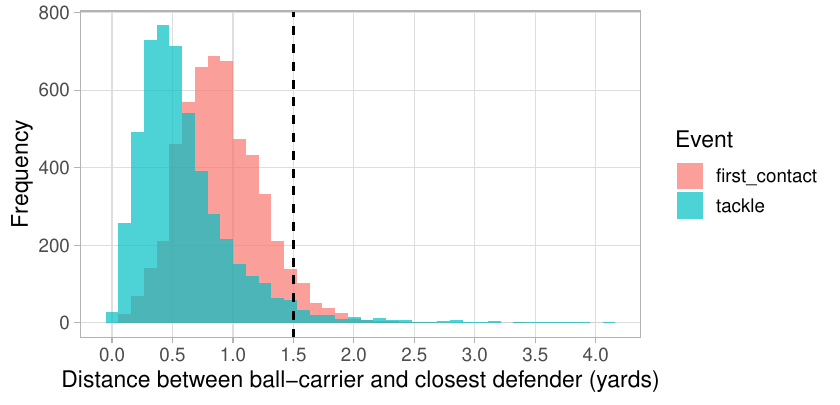} 
\caption{Distributions of the distance between the ball-carrier and closest defender at first contact and tackle events. The dashed line represents the distance value of 1.5 yards, which covers around 95\% of the distances in both distributions. We define a contact window based on consecutive frames that the defensive player remains within this 1.5 yards threshold of the ball-carrier during a play.}\label{fig:hist_dist_bc}
\end{figure}

We define a contact window as a set of consecutive frames within a play where at least one defender is within a distance threshold $D$ (yards) of the ball-carrier.
We first observe the distribution of the distance between the ball-carrier and nearest defender at the moment of first contact event across all considered run plays.
Using this distribution, we then pick a distance threshold $D$ that covers a sufficiently large fraction of tackle events.
This means that when a tackle event is recorded in the tracking data, we also say that the defender is in contact with the ball-carrier.
We recognize that the annotated first contact event is likely not perfect or at the exact moment of contact, but it is the only label provided in the tracking data beyond the designation of tackles.
Figure \ref{fig:hist_dist_bc} shows the distribution of the distance between the ball-carrier and closest defender for first contact and tackle events observed from the tracking data.
The dashed line indicates our chosen threshold of $D=1.5$, which captures about 95\% of the distances across both first contact and tackle events.

\begin{figure}[t]
\centering
\includegraphics[width=1\linewidth]{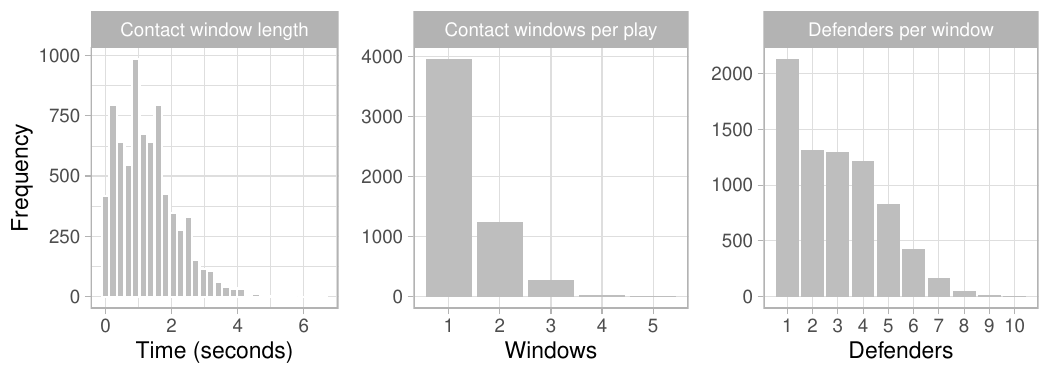} 
\caption{
\textit{Left:} Distribution for the duration of contact windows.
\textit{Center:} Counts of contact windows per play.
\textit{Right:} Counts of defensive players per contact window.
}
\label{fig:window_summary}
\end{figure}

Across the 5,539 considered run plays, we identify 7,453 total contact windows.
Figure \ref{fig:window_summary} presents visual summaries for the contact windows.
The window length appears to be short in general, with an average of 1.28 seconds.
The number of contact windows per play ranges between 1 and 5, with 1 having the highest count.
In most cases, one single defender is in the contact window of the ball-carrier.
Notice that multiple defenders can also be in the contact window of the ball-carrier.
This indicates shared credit among the defenders on stopping the ball-carrier, which motivates a procedure for attributing credit among the defenders involved in each contact window.

\subsubsection{Valuing a contact window}

Let $v_{ijt}$ be the velocity toward the end zone  of the ball-carrier at frame $t = 1, \dots, T_{ij}$ of contact window $j = 1, \dots, J_i$ for play $i = 1, \dots, I$.
Let $v_{ij}^\text{start} = v_{ij1}$ and $v_{ij}^\text{end} = v_{ijT_{ij}}$ represent the velocity toward the end zone for the ball-carrier at the starting and ending frames of contact window $j$ of play $i$, respectively.
Let $v_{ij}^\text{pre}$ denote the peak (maximum) velocity achieved by the ball-carrier either before or during contact window $j$ of play $i$.
For a contact window, we measure the change in velocity between its starting and ending frames relative to the peak velocity achieved by the ball-carrier across the entire play.
Formally, the value for contact window $j$ within play $i$ is determined as
\begin{equation}\label{eq:value}
w_{ij} = \displaystyle \frac{v_{ij}^\text{start} - v_{ij}^\text{end}}{v_{ij}^\text{pre}}.
\end{equation} 
We modify Equation \ref{eq:value} given different cases that may arise:
\begin{itemize}
\item
  If $v_{ij}^\text{pre}$ happens within contact window $j$ of play $i$, we replace $v_{ij}^\text{start}$ with $v_{ij}^\text{pre}$ in Equation \ref{eq:value}. This ensures that we do not penalize defenders for suppressing the peak ball-carrier velocity and maintaining a consistent peak across contact windows.
\item
  Let $v_{ij}^\text{post}$ denote the maximum observed velocity after contact window $j$ of play $i$.

  \begin{itemize}
  \item
    If $v_{ij}^\text{post} \geq v_{ij}^\text{pre}$, we set $w_{ij} = 0$, since the ball-carrier recovers all of their forward motion to essentially begin a new run.
  \item
    If $v_{ij}^\text{end} \leq v_{ij}^\text{post} < v_{ij}^\text{pre}$, we replace $v_{ij}^\text{end}$ with $v_{ij}^\text{post}$ Equation \ref{eq:value}. This ensures that the contact window value represents the fraction of unrecovered peak velocity.
  \end{itemize}
\end{itemize}

\subsubsection{Crediting individual players}

For defender $k = 1, \dots, K_{ijt}$ involved at frame $t$ of contact window $j$ during play $i$, we divide the contact window value $w_{ij}$ for each defensive player as follows.
\begin{itemize}
\item
  \emph{(Individual frame credit).} First, we distribute the window-level value $w_{ij}$ equally across all frames within the window. Thus, the value for frame $t$ of contact window $j$ during play $i$ is $$w_{ijt} =\frac{w_{ij}}{T_{ij}},$$ where $T_{ij}$ is the length of contact window $j$ of play $i$.
\item
  \emph{(Frame-level player credit).} We then divide the frame-level value $w_{ijt}$ equally across all $K_{ijt}$ involved defenders to obtain the credit for defender $k$ at frame $t$ within contact window $j$ of play $i$. Formally, $$w_{ijkt} = \frac{w_{ijt}}{K_{ijt}}.$$ We call these credited portions ``fractional tackles'' for each player.
\item
  \emph{(Window-level player credit).}  For each contact window, we obtain the fractional tackles for each defender across the entire window by tallying up their credited portions across their involved frames. Thus, the fractional tackles recorded by defender $k$ in contact window $j$ of play $i$ is determined as $$w_{ijk} = \sum_{t=1}^{T_{ij}} w_{ijt}.$$
\end{itemize}

\section{Results}
\label{sec:results}

\subsection{Illustration of our framework}
\label{sec:illustration}

\begin{figure}[t]
\centering
\includegraphics[width=0.9\linewidth]{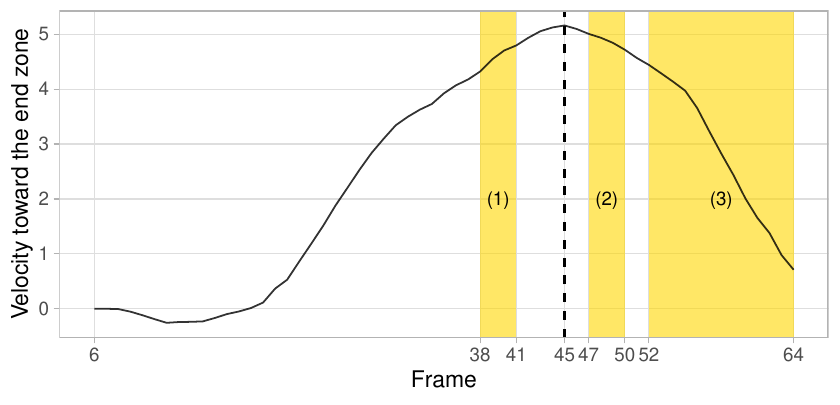} 
\caption{
The change in velocity toward the end zone of the ball-carrier during the seven-yard run by Saquon Barkley.
The shaded regions indicates the three identified contact windows for this play.
}
\label{fig:velo_curve}
\end{figure}

To illustrate our framework, we use the example run play by Saquon Barkley as mentioned in Section \ref{sec:tracking}.
Figure \ref{fig:velo_curve} is a line graph showing how the ball-carrier's velocity in the end zone direction changes throughout the play, with the shaded regions indicating the contact windows.
For this play, we observe three contact windows.
We now demonstrate the latter two steps of our framework as described in Section \ref{sec:framework}.

First, for contact window (1), the peak velocity occurs after the window (see Figure \ref{fig:velo_curve}), as the ball-carrier continues to gain velocity toward the end zone after this window.
Thus, window (1) and the involved defenders receive no credit.

\begin{figure}[t]
\centering
\includegraphics[width=0.95\linewidth]{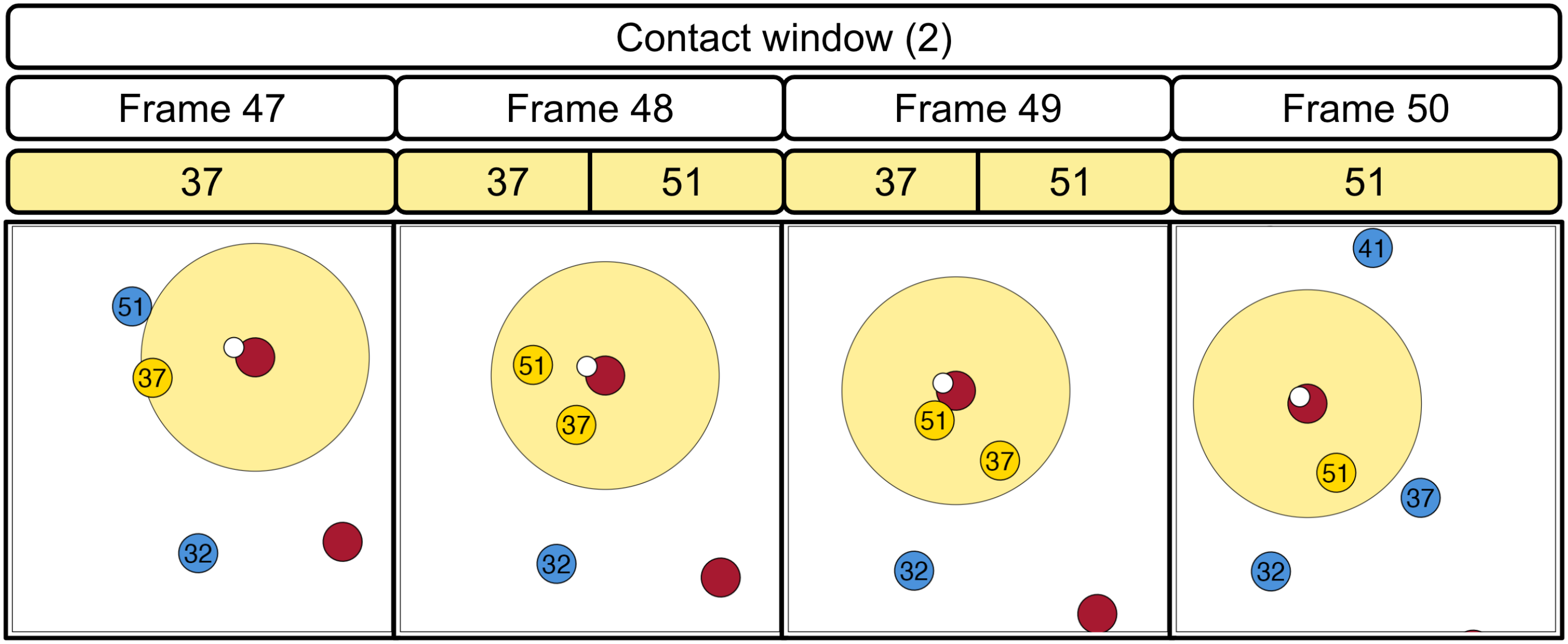}
\caption{
Our procedure for dividing contact window value across all frames and defensive players involved, illustrated with contact window (2) of the seven-yard run by Saquon Barkley.
The red and blue circles represent the offense (New York Giants) and defense (Tennessee Titans), respectively.
Defenders within a radius of 1.5 yards of the ball-carrier at each frame are highlighted in gold.
In this procedure, the window-level value is divided equally across all frames; then the frame-level value is divided equally across all defenders (represented by their jersey numbers) in contact with the ball-carrier at each frame.}\label{fig:credit}
\end{figure}

Next, we focus on contact window (2) as our main illustration.
We notice that the ball-carrier's velocity toward the end zone reaches its peak before window (2) begins (see Figure \ref{fig:velo_curve}).
Figure \ref{fig:credit} illustrates the credit division procedure at the window, frame, and player-level for window (2), along with the locations on the field of Giants (in red, on offense) and Titans (in blue, on defense) players at each frame.
Note that the Titans defenders are color-coded gold when they are in the contact window of the ball-carrier, which are the circular regions around the ball-carrier at each frame in Figure \ref{fig:credit}.
For context, contact window (2) spans four frames within the play: 47, 48, 49, and 50.
There are two defenders involved in the ball-carrier's contact window---Amani Hooker (number 37) at frames 47, 48, and 49; and David Long (number 51) at frames 48, 49, and 50.
Contact window (2) starts at frame 47 with $v_{2}^\text{start} = 5.01$ yards/second and ends at frame 50 with $v_{2}^\text{end} = 4.72$ yards/second.
The peak velocity prior to window (2) is $v_{2}^\text{pre} = 5.16$ yards/second at frame 45 (see Figure \ref{fig:velo_curve}).

To obtain the window-level value, we apply Equation \ref{eq:value} and get 0.056 as the value for contact window (2).
Next, we distribute this value equally across the four frames of window (2).
Thus, each frame receives a value of 0.014.
For each frame, we divide the frame-level value equally across all players that are in the ball-carrier's contact window, based on how many players there are.
This gives us the frame-level fractional tackles for each player.
For frame 47 with one player involved (Amani Hooker), he receives the full frame-level value of 0.014 fractional tackles.
For frames 48 and 49 with two players overlapping, the frame-level credit gets split in half and each player receives 0.007 fractional tackles.
Similar to frame 47, since frame 50 only has one player in the ball carrier's contact window (David Long), he receives 0.014 fractional tackles.
In total, Hooker and Long both receive 0.028 as their respective fractional tackles across contact window (2).

For brevity, the credit attribution for defenders involved in contact window (3) proceeds similarly as for contact window (2).

Table \ref{tab:ftsummary} summarizes our fractional tackles metric for the Tennessee Titans defenders in the Saquon Barkley run play.
Previously, Zach Cunningham gets the tackle credit on the play by the NFL, with an assist by Ugochukwu Amadi.
Here, we are able to credit four additional defenders for halting the ball-carrier's forward motion, which effectively downweights the contributions of those originally recorded a tackle or assist.
In addition, Pro Football Focus records a missed tackle for David Long during this play.
Despite the missed tackle, we still credit Long for contributing to slowing down the
ball-carrier.

\begin{table}[t]
\caption{Summary of fractional tackles for the Tennessee Titans defenders in the seven-yard run by Saquon Barkley. \label{tab:ftsummary}}
\centering
\begin{tabular}{llccl}
\hline
Player & Position & Window & Fractional Tackles & Play-Level Statistic \\ 
\hline
Bud Dupree & OLB & 1 & 0.000 & \\
Amani Hooker & FS & 2 & 0.028 & \\
David Long & ILB & 2 & 0.028 & Missed tackle \\
Denico Autry & DE & 3 & 0.056 & \\
Jeffery Simmons & DT & 3 & 0.130 & \\
Ugochukwu Amadi & FS & 3 & 0.269 & Assist \\
Zach Cunningham & ILB & 3 & 0.269 & Tackle \\
\hline
\end{tabular}
\end{table}

\subsection{Linebackers and linemen dominate fractional tackles leaderboard}

For each defender, we obtain the total and average fractional tackles aggregated across all run plays by running backs from the first nine weeks of the 2022 season.
Table \ref{tab:leaders} shows the top-performing NFL defenders rated by fractional tackles.
Our results make intuitive sense and align with conventional tackles rankings.
Specifically, we find consistency between our list and the \href{https://www.nfl.com/stats/player-stats/category/tackles/2022/reg/all/defensivecombinetackles/desc}{2022 NFL tackles leaderboard}, with Roquan Smith, Foyesade Oluokun, and Zaire Franklin finishing in the top six both according to our metric and in reality.

In general, the top defenders based on total fractional tackles are linebackers (ILBs and OLBs).
However, we also observe defensive linemen (DTs, DEs, and NTs) throughout the leaderboard with notably higher average fractional tackles across all considered run plays relative to other positions.
Thus, although linebackers accumulate tackles, our approach captures how defensive linemen accumulate consistent stops on the ball-carrier's forward motion.
This highlights how our fractional tackles metric provides a measurable statistic for defensive linemen, a position that is sufficiently lacking objective metrics.

\begin{table}
\caption{Fractional tackles leaders across all run plays by running backs during weeks 1 to 9 of the 2022 NFL regular season. \label{tab:leaders}}
\centering
\begin{tabular}{llcccc}
\hline
Player & Position & Plays & Windows & \begin{tabular}[c]{@{}c@{}}Total\vspace{-1.5mm}\\Fractional Tackles\end{tabular} & \begin{tabular}[c]{@{}c@{}}Average\vspace{-1.5mm}\\ Fractional Tackles\end{tabular} \\
\hline
Roquan Smith & ILB & 195 & 107 & 19.83 & 0.102 \\ 
C.J. Mosley & ILB & 184 & 103 & 18.19 & 0.099 \\ 
Christian Wilkins & DT & 140 & 80 & 17.53 & 0.125 \\ 
Divine Deablo & OLB & 146 & 81 & 17.35 & 0.119 \\ 
Foyesade Oluokun & ILB & 181 & 104 & 16.73 & 0.092 \\ 
Zaire Franklin & OLB & 207 & 109 & 16.63 & 0.080 \\ 
DeForest Buckner & DT & 141 & 71 & 16.55 & 0.117 \\ 
Grover Stewart & DT & 161 & 85 & 16.52 & 0.103 \\ 
Jordyn Brooks & ILB & 185 & 97 & 16.48 & 0.089 \\ 
Drue Tranquill & ILB & 162 & 95 & 16.44 & 0.101 \\ 
Rashaan Evans & ILB & 182 & 102 & 16.41 & 0.090 \\ 
Bobby Okereke & ILB & 175 & 93 & 16.07 & 0.092 \\ 
Derrick Brown & DE & 134 & 78 & 16.06 & 0.120 \\ 
Alex Anzalone & ILB & 160 & 86 & 16.00 & 0.100 \\ 
Devin White & ILB & 190 & 87 & 15.82 & 0.083 \\
\hline
\end{tabular}
\end{table}

\subsection{Solo and assisted tackles are overstated statistics}

\begin{figure}[t]
\centering
\includegraphics[width=0.8\linewidth]{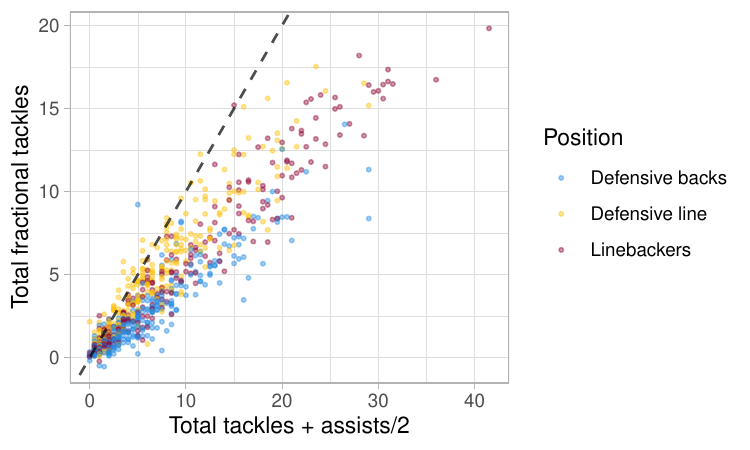}
\caption{
Relationship between total fractional tackles and \amend{traditional counting statistics of total combined tackles (tackles + assists/2)} aggregated across all run plays by NFL running backs over the first nine weeks of the 2022 NFL season.
Despite the strong correlation ($r=0.93$), total fractional tackles are clearly lower than total combined tackles, suggesting that the discrete outcomes of tackles and assists overstate a defender’s tackling contribution.
}
\label{fig:overstate}
\end{figure}

Next, we validate our proposed fractional tackles metric by assessing its statistical properties.
Here, we compare our metric with a simple measure of combined tackles, defined as total tackles and assists/2 (i.e., each assist counts as 0.5).
Figure \ref{fig:overstate} displays a scatterplot of total fractional tackles and total combined tackles for NFL defenders across all considered run plays.
In this graph, the dashed line is the identity line, which means if the total fractional tackles for a player exactly equals their total combined tackles, their corresponding data point would fall on this line.

We observe a strong positive correlation ($r=0.93$, \amend{95\% CI: $(0.92, 0.94)$}), indicating that fractional tackles are generally aligned with recorded tackles.
However, as reflected by the identity line, we see substantially lower values for total fractional tackles in comparison to tackles and assists.
This is likely because the conventional discrete statistics overstate a player's contribution to halting ball-carrier's forward motion.
We also notice clear variation between players with the same number of tackles and assists.
In particular, defensive linemen tend to display higher fractional tackles than other positions with the same number of combined tackles.
This makes intuitive sense, as defensive linemen in run plays often have the first opportunity to interfere with the ball-carrier's forward motion.

\subsection{Fractional tackles offer greater stability than the traditional counting statistics}

We now investigate the stability \amend{(i.e., reliability)} of fractional tackles over the first nine weeks of the 2022 NFL regular season.
To do so, we split our data into two time periods---first four weeks and last five weeks, and examine whether past performance is correlated with future performance.
This gives us an indication of whether our statistic measures the same quantity over time, which is a desired statistical property of a good performance metric \citep{franks2016meta}.

\begin{figure}[!h]
\centering
\includegraphics[width=0.99\linewidth]{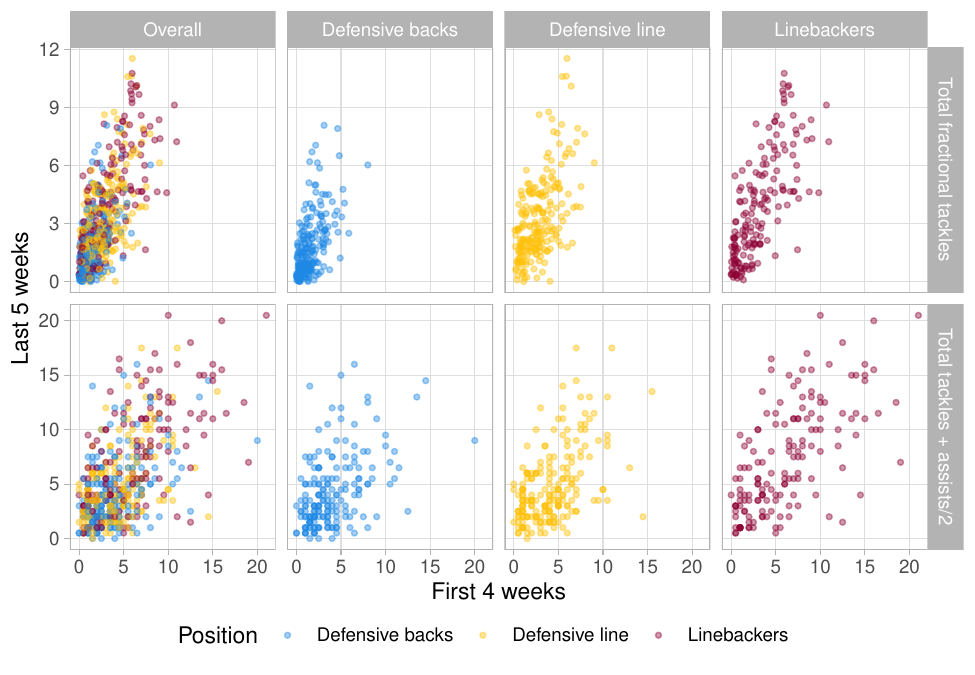} 
\caption{
Overall and within-position relationships for total fractional tackles \amend{(our proposed metric, top row)} and total combined tackles \amend{(traditional counting statistics, bottom row)} aggregated across all run plays by NFL running backs from the first four weeks versus last five weeks of the nine-week data.
Fractional tackles are more predictive of future performance than the traditional counts of tackles and assists.
}
\label{fig:stable}
\end{figure}

In Figure \ref{fig:stable}, the top row displays scatterplots of total fractional tackles for the first four weeks against the last five weeks of our data.
Likewise, the bottom row contains the same set of graphs but for the \amend{traditional counting statistics of combined tackles (i.e., tackles + assists/2)}, for the purpose of comparison.
In each row, the first plot shows the overall relationship whereas the latter three depict the correlation after residualizing for defensive position (defensive backs, defensive line, and linebackers).
\amend{
Additionally, Table \ref{tab:corr} accompanies Figure \ref{fig:stable} with the correlation value for each relationship along with the corresponding 95\% confidence interval.
}


We observe a moderately strong correlation ($r=0.69$, \amend{95\% CI: $(0.65, 0.73)$}) in fractional tackles between the two time periods (see the top left plot of Figure \ref{fig:stable}). 
Thus, fractional tackles are repeatable among defenders throughout the first nine weeks of the 2022 season.
Fractional tackles also possesses greater stability than combined tackles and assists, both overall and within-position, as the linear relationship between the first four and last five weeks appears to be stronger across the board for our metric than for the traditional counting statistics.
Hence, our proposed metric is a better indicator of future performance than the conventional binary outcomes of tackles and assists.
This demonstrates that fractional tackles provide great predictive power for capturing defensive ability of NFL players.

\begin{table}[!h]
\caption{
\amend{Pearson correlation coefficients along with 95\% confidence intervals for the overall and within-position relationships shown in Figure \ref{fig:stable}}. \label{tab:corr}}
\centering
\begin{tabular}{ccccl}
\hline
Overall & Defensive backs & Defensive line & Linebackers & \\
\hline
\begin{tabular}[c]{@{}c@{}}0.69\\(0.65, 0.73)\end{tabular} &
\begin{tabular}[c]{@{}c@{}}0.57\\(0.47, 0.65)\end{tabular} &
\begin{tabular}[c]{@{}c@{}}0.57\\(0.48, 0.66)\end{tabular} &
\begin{tabular}[c]{@{}c@{}}0.73\\(0.66, 0.80)\end{tabular} &
Total fractional tackles\\
\begin{tabular}[c]{@{}c@{}}0.59\\(0.54, 0.65)\end{tabular} &
\begin{tabular}[c]{@{}c@{}}0.46\\(0.34, 0.57)\end{tabular} &
\begin{tabular}[c]{@{}c@{}}0.51\\(0.40, 0.61)\end{tabular} &
\begin{tabular}[c]{@{}c@{}}0.64\\(0.54, 0.73)\end{tabular} &
Total tackles + assists/2\\
\hline
\end{tabular}
\end{table}

\section{Discussion}
\label{sec:discussion}

In this paper, we have introduced fractional tackles, a new measure for valuing defensive performance in American football, moving beyond the observed tackle outcome.
Although we specifically focus on run plays by running backs, our framework can be extended to other situations such as quarterback runs or pass plays.
We note that our quantity can be easily implemented in the media during a live broadcast, or in any data pipeline since it is model-free.
\amend{Due to its continuous nature, our metric provides an evaluation of tackling within a play, as opposed to just using the traditional play-level summaries of tackles and assists.
With fractional tackles, we are able to assess more defenders for slowing down the ball-carrier who previously are not credited with the discrete, binary statistics of tackles and assists.
Specifically, across the considered RB run plays over the first nine weeks of the 2022 NFL season, our metric credits 19,691 unique players, which is more than doubled the number of defenders recording either a solo or assisted tackle at the box-score level (7,720 players).}

We also demonstrate that total fractional tackles is highly correlated with traditional tackles and assists, but provides greater variation and measurable information for players lacking recorded statistics like defensive linemen.
\amend{To clarify, although it is intuitive that our metric is well-aligned with total tackles and assists (e.g., linebackers accumulate the highest amount of tackling credits), our goal is not to predict whether the play outcome is a solo or assisted tackle. Instead, compared to the traditional box-score statistics, we are able to assign more fractional tackling credit to defenders over the course of a play for their contribution to halting the ball-carrier's forward motion (i.e., how tackling is officially defined in the NFL Rulebook, as mentioned in Section \ref{sec:introduction}).}

Despite the novel contribution, our approach is not without limitations.
To identify contact windows, we use the annotated first contact event during each play provided by the player tracking data to determine a distance threshold (1.5 yards) within which a defender is in the contact window of the ball-carrier.
While we can perform sensitivity analysis to assess this cutoff, we do not know if the players are actually making contact with one another based on the provided 2D tracking data.
Yet, the contact window definition can be improved with better data (e.g., pose data generated via computer vision) where contact could be explicitly identified.
\amend{
We also recognize that our contact window definition is an empirically-based approach based on only data from the first nine weeks of the 2022 NFL season.
However, this definition can certainly change given that a longer period (e.g., a full season or multiple seasons) of tracking data is available.
Additionally, our current definition for contact windows is player-agnostic, but one could theoretically assume that these windows vary by player-specific traits (e.g., arm span), prompting an avenue for future research.
Besides, by design, it is straightforward to modify the contact window identification step in our flexible framework by replacing our model-free definition with model-based approaches such as those presented by other NFL Big Data Bowl entries \citep{mackenzie2024how}.}

\amend{
In addition, our current framework distributes equal credit across all frames within a contact window, and also gives equal credit to multiple defensive players that are in contact with the ball-carrier at each frame.
We recognize that these are simplifying assumptions, as an ideal method would account for frame-level tracking data to infer attribution weights.  
Furthermore, our analysis is subject to the selection bias regarding a player's involvement in a contact window.
From a defensive standpoint, a defender might be in a contact window because they are being targeted by the offense.
As for the offensive side, it might be the case that a ball-carrier enters a window due to poor decision-making.
We leave the exploration of these challenging issues for the future, which will require careful modeling of frame-level player tracking data that accounts for various space-time, contextual, and on-field personnel features, such as ``ghosting'' techniques for modeling baseline-level player behavior \citep{yurko2024nfl}.
}

In general, note that the fractional tackle values across all windows in a play do not add up to one (see, for example, the results summarized in Table \ref{tab:ftsummary}).
This can be due to reduction in velocity outside of contact windows or if the play ends while the ball-carrier has positive velocity (e.g., running out of bounds, being down by contact but the body is still motioning forward, etc.), but we leave this exploration for future work.
Additionally, for plays where a ball-carrier recovers the velocity (and effectively begins a new run), the current method does not credit defenders for what happens before the ball-carrier's peak velocity toward the end zone. 
To address this issue, it will be necessary to explore the temporal dependence between contact windows in the future. 

Furthermore, we believe there are other relevant player movement attributes for gaining insights on tackling in American football.
In this paper, we consider the velocity in the direction of the target end zone as the measure for the forward motion of a ball-carrier.
However, ball-carriers differ in size; thus tackling a powerful 240-pound running back is very different from tackling one who is much lighter in weight.
To address this, we could consider the concept of momentum, which represents the idea of ``mass in motion''.
Formally, the momentum of an object is defined as the product of its mass and velocity.
In our football setting, we could obtain a ball-carrier's momentum toward the end zone at every moment within a play directly from the rich player tracking data.
This would provide a more reasonable measure of forward motion than only using velocity, since the physical size of a player is also accounted for.
Using similar ideas to what we have proposed in this work, we could examine the observed change in momentum of a ball-carrier in future studies on tackling in American football.
We note that replacing velocity with momentum in our credit assignment framework to create fractional tackles would yield the same results, since the change in contact window value relative to the peak momentum or velocity are equivalent due to constant ball-carrier weight throughout the play.
Still, there are potential opportunities for exploring which features matter most in disrupting the momentum toward the end zone of a ball-carrier (e.g., angle of attack, player orientation, defender momentum, etc.), along with investigating changes in ball-carrier momentum caused by
defensive action that is not necessarily within a contact window (e.g., when the ball-carrier notices a path is blocked and has to adjust). 

Nevertheless, we believe our contribution of fractional tackles is a significant step forward in evaluating defensive contribution in American football, as well as a clear example of what the complex player tracking data can offer to the task of measuring player performance in sports.

\section*{Acknowledgments}

We thank the organizers of the NFL Big Data Bowl 2024 for hosting the
competition and providing access to the data. Q.N. thanks Gregory J.
Matthews for a conversation ``at work'' that inspired the conceptualization of this
work.

\section*{Code and data availability}

The data provided by the NFL Big Data Bowl 2024 are available at \url{https://www.kaggle.com/competitions/nfl-big-data-bowl-2024/data}.
Our results can be reproduced using the scripts provided at \url{https://github.com/qntkhvn/tackle}.

\bibliography{references}

\end{document}